\documentclass[prd,nofootinbib,twocolumn]{revtex4}
\usepackage{graphicx,natbib}
\usepackage{url}
\usepackage[usenames]{color}
\usepackage{rotating}
\usepackage{multirow}
\usepackage{amssymb,amsmath}
\usepackage{psfrag}
\usepackage{bm}
\usepackage{hyperref}

\newcommand{\Ngal}{N_{\rm gal}}
\newcommand{\Mobs}{M_{\rm obs}}
\newcommand{\Mbias}{M_{\rm bias}}
\newcommand{\MbiasBlend}{M_{\rm bias,b}}

\newcommand{\Mth}{M_{\rm th}}
\newcommand{\Msun}{M_{\odot}}

\newcommand{\siglnM}{\sigma_{{\rm ln} M}}
\newcommand{\siglnMBlend}{\sigma_{{\rm ln} M,{\rm b}}}
\newcommand{\sigmaBlend}{\sigma_{0,{\rm b}}}

\newcommand{\DE}{\Omega_{\rm DE}}

\newcommand{\eg}{\hbox{\sl e.g.,\ }}
\newcommand{\hinv}{\hbox{\, $h^{-1}$ }}
\newcommand{\mpc}{\hbox{\, \rm Mpc}}
\newcommand{\Planck}{\hbox{\sl Planck}}
\newcommand{\maxBCG}{\hbox{\tt maxBCG}}
\newcommand{\maxbcg}{\hbox{\tt maxBCG}}
\newcommand*{\thetaCosm}{\ensuremath{\bm{ \theta}}}

\def\spose#1{\hbox to 0pt{#1\hss}}
\def\lta{\mathrel{\spose{\lower 3pt\hbox{$\mathchar"218$}}
     \raise 2.0pt\hbox{$\mathchar"13C$}}}
\def\gta{\mathrel{\spose{\lower 3pt\hbox{$\mathchar"218$}}
     \raise 2.0pt\hbox{$\mathchar"13E$}}}

\begin{document}

\title{Influence of Projection in Cluster Cosmology Studies}
\author{Brandon~M.~S.~Erickson$^{1}$,Carlos~E.~Cunha$^{1}$ and August~E.~Evrard$^{1,2,3}$}
\affiliation{$^1$Physics Department, University of Michigan, Ann Arbor, MI 48109, USA}
\affiliation{$^2$Astronomy Department, University of Michigan, Ann Arbor, MI 48109, USA}
\affiliation{$^3$Michigan Center for Theoretical Physics, Ann Arbor, MI 48109, USA}

\date{\today}

\begin{abstract}
Projection tends to skew the mass--observable relation of galaxy clusters by creating a small fraction of severely blended systems, those for which the measured observable property of a cluster is strongly boosted relative to the value of its primary host halo.  We examine the bias in cosmological parameter estimates caused by incorrectly assuming a  Gaussian (projection-free) mass--observable relation when the true relation is non-Gaussian due to projection.   We introduce a mixture model for projection and explore Fisher forecasts for a survey of 5000 deg$^2$ to $z=1.1$ and an equivalent mass threshold of $10^{13.7} \, h^{-1} \Msun$.  Using a blended fraction motivated by optical cluster finding applied to the Millennium Simulation, and applying \Planck\  and otherwise weak priors, we find that the biases in $\DE$ and $w$ are significant, being factors of $2.8$ and $2.4$, respectively, times previous forecast uncertainties.   Incorporating eight new degrees of freedom to describe cluster selection with projection increases the forecast uncertainty in $\DE$ and $w$ by similar factors.  Knowledge of these additional parameters at the $5\%$ level limits degradation in dark energy constraints to $\lta10\%$ relative to projection-free forecasts.   We discuss strategies for using simulations and complementary observations  to characterize the fraction of blended clusters and their mass selection properties.
\end{abstract}

\maketitle

\section{Introduction}\label{sec:intro}

Galaxy clusters offer tests of large-scale gravity and cosmology, as their space density is exponentially sensitive to the time-dependent amplitude of the matter power spectrum and the cosmic expansion history  (see \citet{Allen1103.4829}  and \citet{Voit0410173} for recent reviews).  Because their counts and clustering probe the gravitational growth of structure, clusters provide information beyond that provided by CMB and cosmic distance measurements such as supernovae and baryon acoustic oscillations.  Joining cluster data  with such measurements significantly improves cosmological parameter constraints \cite{cun09}. 

While the potential for clusters to constrain parameters such as the dark energy equation of state, the energy densities of cosmic components, and the amplitude of matter density fluctuations has long been known \citep{Henry91, Bahcall9803277, Weller0110353, Majumdar0305341, lim04}, early work also emphasized the importance of understanding systematic errors associated with survey modeling \citep[\eg][]{Levine0204273, Evrard0110246}.   
The agreement in cosmological parameters derived recently from independent samples selected at optical \citep{Rozo0902.3702} and X-ray \citep{Henry0809.3832, Mantz0909.3098,  Vikhlinin0812.2720} wavelengths indicates progress in addressing systematic errors.  However,  early \Planck\ analysis of the thermal Sunyaev-Zel'dovich (SZ) effect in the optically-selected \maxBCG\  sample \citep{Koester0701265}  suggests more work to be done; the mean gas thermal energies inferred from 
\Planck\  measurements lie a factor of two below simple model expectations \citep{Planckoptical11}.  While the origins of this disagreement are not yet fully known, the effective offset in total mass of $\sim 40\%$ is in $\sim 2.5 \sigma$ conflict with the mass calibration errors quoted in the above cosmological studies.  The tension may be partly relieved by introducing a $\sim 20\%$ offset in X-ray and weak lensing masses \cite{Meneghetti0912.1343}, but other effects need to conspire and amplify this.  

The \Planck\  optical results highlight the importance of a key functional element of cluster cosmology from counts, namely the likelihood, $p(\Mobs | M,z)$, that a halo of mass $M$ at redshift $z$ has an observed property $\Mobs$.  For SZ observations, $\Mobs$ is the total thermal energy of the hot gas inferred from the spectral distortion in the cosmic microwave background.  For the case of optical studies we consider $\Mobs$ to be the optical richness, defined by the number of red galaxies in the cluster above a given magnitude limit.   Sky counts expected within a particular model are calculated by a convolution of this mass--observable function with the space density of halos.  The latter has been calibrated to high precision by N-body simulations \cite{Jenkins0005260, Evrard0110246, Warren0506395, Tinker0803.2706, Crocce0907.0019}.  

Since the scaling of most observables with mass are power-laws, and therefore linear in the logarithm, the convolution kernel is typically assumed to be log-normally distributed  about a power-law mean.  
The log-normal assumption for deviations in hot gas properties about the mean is supported by X-ray observations of core-excised luminosity and temperature in clusters \cite{Pratt0809.3784} and from a range of hot gas properties in simulated halo samples  \cite{Stanek0910.1599, Yang0808.4099}.   However, some degree of departure from log-normality should be expected intrinsically, potentially driven by different formation histories \cite{Yang1010.0249} and by major merging events \cite{Poole0608560}.   

A given intrinsic likelihood for halo observables will be modified when $\Mobs$ is projected onto the sky \cite{White0205437, coh07, Hallman0704.2607}.  
Halos projected along the line-of-sight of a given target boost its signal in a stochastic manner, resulting in a $P(\Mobs | M,z)$ that grows a tail to high values.   Optical richness, an attractive property to measure because it only requires broad-band photometry, is sensitive to line of sight projections.  Richness scales roughly linearly with mass \cite{Johnston07, Zheng0809.1868, White1010.4915} while X-ray and SZ signals scale more steeply, $\sim M^{1.6}$ \cite{Reiprich0111285, Stanek0602324, Rykoff0802.1069, Planckxraystack11}, making these observables less susceptible to contamination from (spatially more abundant) lower mass halos.  Indeed, the Abell catalog \cite{abell58, abell89} contains cautionary notes about projected confusion.  Spectroscopic studies of optically selected clusters occasionally reveal multiple peaks and complex structures in velocity space (A1689 \cite{Lokas06}, CL1604 \cite{gal08}, A85 \cite{boue08064262} and EIS clusters \cite{Grove0809.4552}), and simulation studies are beginning to explore these issues in detail \cite{White1005.3022, Noh1011.1000}.   Joint X-ray and optical studies of three nearby clusters show thermal signatures anticipated by gas dynamic simulations for actively merging systems\cite{Maurogordato1009.1967}.  

The statistical ingredients (the space density, spatial clustering, and galactic content of halos) needed to calculate projected confusion are coming into focus, and a generic expectation is that most massive halos suffer little contamination while a modest percentage are strongly affected by projection \cite{coh07, Rozo1104.2090}.   These studies motivate a Gaussian mixture model for projection that we explore in this paper.   The mixture represents a dominant component of clusters whose sightlines are largely clean along with a minority of clusters whose signal is strongly boosted.   The latter category we refer to as blended systems, or {\sl blends}, and in these objects the $\Mobs$ signal is not dominated by a single halo.  Our treatment here is intended to be illustrative, but model parameters could be tuned using sky simulations tailored to specific surveys \citep{Cai0810.2300}.  

An earlier study of projection used an Edgeworth expansion to model cluster counts including non-zero skewness and kurtosis in $p(\Mobs | M,z)$.  \citet{shaw10} find that the detailed shape becomes important when the product of the scatter in the mass--observable relation and the logarithmic slope of the mass function at the limiting mass is greater than one.   Our study differs from that work in two ways:  
our Gaussian mixture approach, which includes eight new degrees of freedom, is more general than their expansion, and we use a Fisher matrix approach to explicitly calculate both the bias that projection induces in a projection-free (single component Gaussian) analysis and the additional variance that is incurred when the extra degrees of freedom are included.  We explore the latter under a variety of prior constraints.  

The paper is organized as follows.  In \S \ref{sec:counts} we briefly recount the procedure for how to extract dark energy constraints by computing cluster counts and variance in counts, and present our parameterization of the mass--observable relation including the line-of-sight projection.  In \S \ref{sec:res}, we present our results and, in \S \ref{sec:conc}, a discussion of the results.


\section{Cluster Selection Model with Projection}\label{sec:selection}

When Abell published the first homogeneous cluster catalog from photographic plate imaging, he employed the count of galaxies within a fixed metric aperture and a scaled magnitude range as a measure of {\sl galactic richness}, a proxy for cluster mass \cite{abell58}.  The development of multi-band imaging cameras in the late 1990s  \cite{Gunn98, Megacam03}  enabled cluster samples to be selected using color selection techniques, whereby counts within a joint magnitude and color (or photometric redshift) range are employed as a mass proxy \cite{Gladders05, Koester0701265, Wen0906.0803, Hao1010.5503, Szabo1011.0249, Adami0910.3827}.   These samples contain up to 69000 clusters extending to $z \sim 1$ across nearly 8000 deg$^2$ of sky.
The next generation of optical and near-infrared surveys---the Dark Energy Survey\footnote{\url{http://www.darkenergysurvey.org/}} (DES), the VISTA surveys\footnote{\url{http://www.eso.org/sci/observing/policies/PublicSurveys}}, Pan-STARRS\footnote{\url{http://pan-starrs.ifa.hawaii.edu/public/}}, with Large Synoptic Survey Telescope\footnote{\url{http://www.lsst.org/lsst}} (LSST) and Euclid\footnote{\url{http://sci.esa.int/euclid}} to follow---will identify hundreds of thousands of clusters. 

Modern, color-based cluster finders rely on the 4000~\AA \ break feature of old stellar populations \cite{bower92}.  Observations show, and stellar population models expect, that the mean color in a fixed observed band straddling 4000~\AA\ will vary with redshift.  A single color can therefore be used as a simple photometric redshift estimator  \cite{Gladders00}.   The redshift accuracy is limited by the finite color width of the red galaxy population at a given epoch.   The finite width of the color filter employed for cluster finding in turn corresponds to a comoving length scale of order hundreds of megaparsecs \cite{coh07}.   Red galaxies in spatially distinct halos that fall within a cylinder of this length aligned toward an observer will be catalogued as a single cluster \cite{milk10}.   We generically refer to this process as {\sl blending}.   While all clusters suffer some degree of projected blending, we are particularly interested in extreme cases, and so adopt a specific definition for classifying clusters.  A cluster of observed richness $\Mobs$ will be referred to as a {\sl blended cluster\/} if no single halo contributes $\Mobs/2$ or more to the richness.   Conversely, a cluster for which a single halo does contribute $ \ge \Mobs/2$ of the richness is referred to as {\sl clean}.  (We assume here that the radial scale for observed and intrinsic measures are aligned.)

The massive halos that host clusters tend to be embedded in filaments and/or supercluster regions.   Viewpaths that traverse such structures will have a locally boosted background in the color-magnitude sub-space used for cluster detection.  Empirical studies of a new red sequence matched filter method applied to SDSS \maxbcg\ data \cite{Rozo1104.2090, Rykoff1104.2089} indicate that such boosts generate a blended fraction of $\sim 10\%$ in the cluster population. 

Existing Fisher matrix forecasts for the cosmological parameter yield from upcoming surveys \cite{Weller0110353, Majumdar0305341, lim04, cun08} have assumed a log-normal distribution for the observable likelihood, $p(\Mobs | M,z)$.   While the log-normal form may reflect the intrinsic (\eg spherically averaged) dispersion in the halo ensemble, blended clusters will have $p(\Mobs|M,z)$ strongly boosted at high $\Mobs$.   We use a Gaussian mixture model in log-mass, described in \S \ref{sec:sysmobs}, to model a bimodal cluster population consisting of clean and blended systems.  

While our model is general, we tune default parameters using the results of \citet{coh07}.  That study applies a red sequence--based algorithm to projected galaxy maps from the Millennium Simulation \citep{Croton06, Springel05}.  They use a single $R-z$ color applied in narrow redshifts sliced centered at $z=0.4$, $0.7$ and $1.0$.   Matching halos to clusters by galactic membership, they identify a blended subset of clusters whose mass--observable relation is shifted to higher $\Mobs$ values and whose variance is larger than that of clean clusters.  At higher redshifts, the mean color in the old stellar population varies more weakly with $z$, and the color width of the red sequence traces out an increasingly longer comoving cylinder, reaching $\sim 500 \hinv\mpc$ at $z=1$.   The longer search cylinder drives an increase in the blended fraction of clusters, from $11\%$ at $z = 0.4$ to 22\% at $z = 1$.  Note that the fraction of halos at fixed true mass that are blended will be lower than this, as convolution with a steeply falling mass function increases the fraction of blended clusters at fixed $\Mobs$ \cite{Rozo1104.2090}.

\subsection{Reference Model Survey}\label{sec:clustermass}

Our reference model survey, based on DES+VISTA\footnote{\tt \url{http://www.vista.ac.uk/}, \url{http://www.darkenergysurvey.org/}}, is assumed to cover 5000 square degrees and extend 
to a limiting redshift of $z_{\rm max} = 1.1$.
Our choice of maximum redshift is somewhat conservative since with the addition of the IR filters from VISTA, the combined surveys should have accurate redshifts for field galaxies up to $z \sim 1.5$.
.  
We assume that DES+VISTA-VHS will detect clusters above an observed threshold, $\Mobs \ge \Mth$, with $\Mth = 10^{13.7} h^{-1} \Msun$, comparable to what is achieved by low redshift surveys \cite{Koester0701265, Johnston07}).  Based on the \maxbcg\ $\Ngal$ richness measure, the zero-redshift variance in the mass--observable relation is taken to be $\sigma_0^2 = 0.25$ \cite{Rozo0809.2794}.

We subdivide the sky into 500 bins of 10 square degrees each, and calculate the 
counts and sample variance using richness bins of width $\Delta_{\ln \Mobs} =0.2$ 
with the exception of the highest mass bin, which we extend to infinity.  
We set the width of our redshift bins to $\Delta z =0.1$.  
These bin sizes imply 11 redshift bins and 10 mass bins.
We assume fiducial cosmological parameters based on the fifth year data release 
of the Wilkinson Microwave Anisotropy Probe (WMAP5, \cite{kom08}).  
Thus, we set the baryon density, $\Omega_b h^2=0.0227$, the dark matter density, 
$\Omega_m h^2 =0.1326$, the normalization of the power spectrum at 
$k=0.05 {\rm Mpc}^{-1}$, $\delta_{\zeta}=4.625 \times 10^{-5}$, the tilt, $n=0.963$, the optical depth to reionization, $\tau=0.087$, the dark energy density, $\DE=0.742$, and the dark energy equation of state, $w=-1$.  In this cosmology, $\sigma_8=0.796$.
With the exception of $w$, the cosmological parameters we use have been determined to an accuracy of a few percent.  
We apply \Planck\  priors\footnote{Planck Fisher matrix courtesy of Wayne Hu.} to all cosmological parameters.
We use CMBfast \citep{sel96}, version 4.5.1, to calculate the transfer functions.

\subsection{The mass--observable relation for clean clusters}\label{sec:sysmobs}

We assume that the majority of clusters are clean systems whose selection properties are described by a single log-normal form.   Following the notation of \cite{cunevr10}, we write the probability of observing a cluster with observable mass proxy, $\Mobs$ given a true mass $M$, as 
\begin{equation}
p(\Mobs | M,z ) = \frac {1}{\sqrt{2\pi \siglnM^2(M,z)}} \exp\left[ -x^2 \right]
\label{eqn:lognormal}
\end{equation}
with
\begin{equation}
x \equiv \frac{ {\rm ln} \Mobs - {\rm ln} M - {\rm ln}  \Mbias(M,z)}{ \sqrt{2 \siglnM^2(M,z)}}.
\end{equation}

The model allows for systematic error in the observable by allowing redshift-dependent bias and variance
\begin{gather}
{\rm ln} \Mbias(M,z) = B_0 + B_1{\rm ln} (1+z), \label{eqn:mbiasc}\\
\siglnM^2(M,z) = \sigma_{0}^2 + \sum_{i=1}^{3} s_i z^i. \label{eqn:msigc}
\end{gather}
We set the fiducial values of  $B_0$, $B_1$ and the $s_i$ to zero throughout this paper.  The baseline mass scatter, $\sigma_0$, is taken to be $0.5$, a value consistent with \maxBCG\ findings for that survey's original $\Ngal$ richness estimator \cite{Rozo0809.2794}.
Recently, \citet{Rykoff1104.2089} proposed an improved mass-estimator for MaxBCG, with scatter expected to be $0.2=0.3$, making our
assumption about the scatter conservative.

Below, we apply this  single Gaussian model to fit a set of data that are described by our extended, two Gaussian case.  For that fit, $\sigma_0$ has a slightly different 
value, and $B_0$, $B_1$ and the $s_i$ elements will be non-zero, as described in \S \ref{ssec:shifts}.

\subsection{Selection with projection: blended clusters}\label{sec:sysmobs2}

To model selection with projection, we use a Gaussian mixture form for $\Mobs$ that combines clean and blended sub-populatons, 
\begin{equation}
\label{eqn:blendgamma}
p(\Mobs | M, z ) =  \left(1 - \gamma(z) \right) \, G_{\rm clean} \, +  \, \gamma(z) \, G_{\rm blend}
\end{equation}
where $G_{\rm clean}$ and $G_{\rm blend}$ are log-normal distributions of the form given by Eq.~\eqref{eqn:lognormal}, and the blend factor, $\gamma(z)$, controls the fraction of blended clusters.  

For the component representing blends, we introduce a set of parameters for the bias and scatter different than that of the clean component,
\begin{gather}
{\rm ln}\MbiasBlend(z) = \mu_0 + \alpha \, {\rm ln}(1+z) + \beta \, ({\rm ln} M - {\rm ln} M_{\rm th}), \label{eqn:mbiasbl}\\
\siglnMBlend^2(M,z) = \sigmaBlend^2 + s_z z + s_M ({\rm ln} M - {\rm ln} M_{\rm th})\label{eqn:msigbl}.
\end{gather}
We highlight below the role of the mass bias terms, especially the constant offset, $\mu_0$, and its logarithmic redshift gradient, $\alpha$.   The parameter $\beta$ allows for a mass-dependent bias.   
For the scatter of the blended component, we focus on a pessimistic scenario where $\sigmaBlend^2 = 2 \sigma^2_{0}$.  This is consistent with results derived from Millennium Simulation analysis \cite{coh07}.  
The more optimistic case of $\sigmaBlend^2 = \sigma^2_{0}$ yields qualitatively similar results.  The variance is allowed to evolve linearly with redshift and log-mass.  

Default parameter values for the blended component model are $\beta = s_z = s_m = 0$, and $\sigmaBlend^2 = 2 \sigma^2_{0}$.  We consider three specific combinations of $\mu_0$ and $\alpha$ that reflect different scenarios of redshift evolution in the bias of the blended component:  none ($\alpha = 0, \ \mu_0 = 0.75$); weak ($\alpha = 0.5, \ \mu_0 = 0.5$); and strong ($\alpha = 1, \ \mu_0 = 0.25$).  In all cases, the log-mean halo mass of blended clusters is biased low, by $\mu_0 + \alpha {\rm ln}(1+z)$  relative that of the clean component.  

The blend factor controls the overall fraction of blended clusters, and we write its evolution as 
\begin{equation}
\gamma(z) = \gamma_0 + \gamma_1 {\rm ln}(1+z) e^{-z}, 
\end{equation}
where the exponential damping is added only to regularize $\gamma$ at high redshift.  The blend factor grows with redshift to $z = 0.77$, then flattens and decreases weakly toward the $z = 1.1$ redshift limit.   

\begin{figure}[!t]
  \begin{minipage}[h]{95mm}
    \begin{center}
      \resizebox{95mm}{!}{\includegraphics{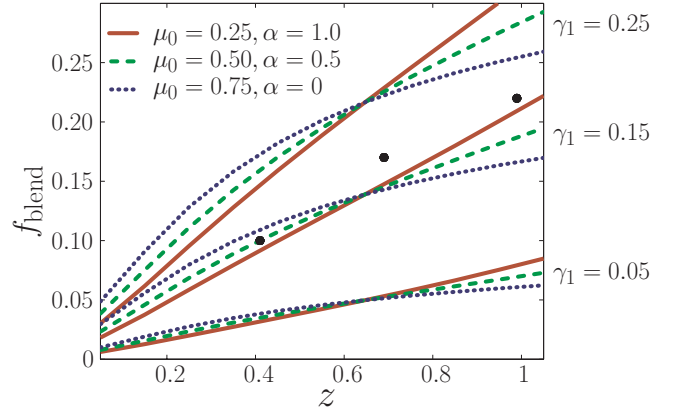}}
    \end{center}
  \end{minipage}
  \caption{
The fraction of blended clusters above the survey $\Mobs$ threshold is 
plotted for $\gamma_0 = 0$ and $\gamma_1 = 0.05,0.15,0.25$ (bottom to top).  
The three black dots are the values found from the Millennium Simulation study of \citet{coh07}. 
Color-styles correspond to three different redshift dependence forms (none, weak, strong) for the bias of the blending amplitude:  $\mu_0 = 0.75, \alpha = 0.0$ (dotted, blue);  $\mu_0 = 0.5, \alpha = 0.5$ (dashed, green); $\mu_0 = 0.25, \alpha = 1$ (solid, red).  The models are tuned to coincide near the median cluster sample redshift of $z=0.65$.  
}
\label{fig:proj}
\end{figure}

We choose this parameterization because it allows sufficient freedom to roughly match the blending fraction as a function of redshift found in \citet{coh07}.   We calculate the blended fraction of clusters as a function of their observable mass proxy,  $\Mobs$, via convolution with the mass function, as described below.
In Fig.~\ref{fig:proj} we show the resulting fraction of blended clusters above the survey threshold,
\begin{equation}
f_{\rm blend} = \frac { \overline{N}_{\rm blend} } { \overline{N}_{\rm blend} + \overline{N}_{\rm clean} }.
\end{equation}
as a function of redshift bin for $\gamma_0 = 0$ and $\gamma_1 = 0.05$, $0.15$ and $0.25$.   The mean counts, $\overline{N}$, are given by Eq.~\eqref{eqn:ccmbar} below, where the clean and blended components are calculated using the associated components of Eq.~\eqref{eqn:blendgamma}.
For each $\gamma_1$, the three lines show results for the three choices of  $\{ \mu_0,\  \alpha \}$ pairs discussed above.  The results of \citet{coh07}, shown as the three black dots in the figure,  are roughly matched by the choice of $\gamma_1 = 0.15$.  

\psfrag{gamma1 = 0.05}{\Huge $\gamma_1 = 0.05$}
\psfrag{gamma1 = 0.15}{\Huge $\gamma_1 = 0.15$}
\psfrag{gamma1 = 0.25}{\Huge $\gamma_1 = 0.25$}
\psfrag{munot = 0.25, alpha = 1.00}{\Huge $\mu_0 = 0.25, \alpha = 1.0$}
\psfrag{munot = 0.50, alpha = 0.50}{\Huge $\mu_0 = 0.50, \alpha = 0.5$}
\psfrag{munot = 0.75, alpha = 0.00}{\Huge $\mu_0 = 0.75, \alpha = 0$}
\psfrag{z}[c][c][1.5][0]{\Huge $z$}
\psfrag{f}[c][c][1.25][0]{\Huge $f_{\rm blend}$}
\psfrag{0}{\Huge $0$}
\psfrag{0.05}{\Huge $0.05$}
\psfrag{0.10}{\Huge $0.10$}
\psfrag{0.15}{\Huge $0.15$}
\psfrag{0.20}{\Huge $0.20$}
\psfrag{0.25}{\Huge $0.25$}
\psfrag{0.2}{\Huge $0.2$}
\psfrag{0.4}{\Huge $0.4$}
\psfrag{0.6}{\Huge $0.6$}
\psfrag{0.8}{\Huge $0.8$}
\psfrag{1}{\Huge $1$}


\subsection{Cluster counts and clustering}\label{sec:counts}

The subject of deriving cosmological constraints from cluster number counts and clustering of clusters has been 
treated extensively in the literature (see, {\sl e.g.}, \cite{cun08,lim04,lim05,lim07}).  
We give a brief summary in this section, following the approach described in \cite{cun08}, 
and leave other details to Appendix \ref{sec:countsandclustering}.

The number density of clusters at a given redshift $z$ with observable in the range $M_a \leq \Mobs \leq M_{a+1}$ is given by

\begin{equation}
\overline n_{a}(z) \equiv \int_{M_a}^{M_{a+1}} \frac{d \Mobs}{\Mobs}
\int {\frac{dM}{M}} { \frac{dn}{d{\rm ln} M}} \,
p(\Mobs | M, z).
\label{eqn:nofz}
\end{equation}
\noindent where $\frac{dn}{d{\rm ln} M}$ is the mean halos space density, also called the mass function.  
We use the Tinker parameterization for the mass function, and ignore errors in redshift estimates.
The mean cluster number counts, and sample covariance, in bins labeled by $i=\{a,b,c\}$, where $a$ denotes mass proxy, $b$ redshift, and $c$ angular coordinate, are given by
\begin{gather}
\label{eqn:ccmbar}
\overline{N}_{i} = \int_{z_b}^{z_{b+1}} dz \, \frac{dV}{dz} \, \overline n_{a} \, W^{\rm th}_c(\Omega) \\
\label{eqn:ccSij}
S_{ij} = \langle (N_{i} -\overline N_{i})(N_{j} - \overline N_{j})\rangle.
\end{gather}
\noindent where ${W_c}^{\rm th}(\Omega)$ is an angular top-hat window function. 

Define the covariance matrix of halo counts
\begin{equation}
C_{ij} = S_{ij} + {\overline N_i} \delta_{ij}
\label{eqn:covdat}
\end{equation}
\noindent where ${\overline N_i}$ is the vector of mean counts defined in Eq.~\eqref{eqn:ccmbar} and $S_{ij}$ is the sample covariance defined in Eq.~\eqref{eqn:ccSij}.  The indices $i$ and $j$ refer to observable, redshift and angular coordinate bins.  Assuming Poisson noise and sample variance are the only sources of noise, the Fisher matrix is, \citep{hu06,lim04,hol01}

\begin{equation}
F_{\alpha\beta}=  \overline{N}^t_{,\alpha} {\bf C}^{-1}
 \overline{N}_{,\beta} 
+ \frac{1}{2} {\rm Tr} \left\{ {\bf C}^{-1} {\bf S}_{,\alpha}
 {{\bf C}}^{-1} {\bf S}_{,\beta} \right\},
 \label{eqn:fishmat}
\end{equation}
where commas denote derivatives with respect to the model parameters.  
The first term on the right-hand side contains the information from 
the mean counts, $\overline N$.  The $S_{ij}$ matrix can be thought of as contributing 
noise to this term, and hence only reduces the information content from counts alone.  
The second term contains the information from the sample covariance.

The marginalized error in a parameter is given by $\sigma(\theta_{\alpha}) = [ (F^{-1})_{\alpha \alpha} ]^{1/2}$.
Priors are easily included in the Fisher matrix.
If parameter $\theta_{i}$ has a prior uncertainty of $\sigma(\theta_{i})$, 
we add $\sigma^{-2}(\theta_{i})$ to the $F_{i i}$ entry of the Fisher matrix before inverting.


\section{Results}\label{sec:res}

Our model with projection differs from previous models that assume an entirely clean (single log-normal) population.  Applying a clean-only model to a sky with projection will generally introduce a bias into derived cosmological parameters.  We first address the magnitude of this bias, then turn to the impact that introducing extra degrees of freedom to represent blends has on marginalized constraints of dark energy parameters.  

\subsection{Parameter Bias}\label{ssec:shifts}

To estimate the bias in cosmological constraints that would result if cluster samples with projection are analyzed using a model with no projection, we follow a linearized approach used in previous studies \cite{huttak05,wu08}.   Our ``true'' sky counts are based on the bimodal mass--observable relation, Eq.~\ref{eqn:blendgamma}, applied using the three redshift evolution cases for the mass bias of the blended component (none, weak, strong) discussed above.  The redshift growth rate of the blending factor, $\gamma_1$, is a controlling degree of freedom.  

If the true sky is analyzed assuming no projection, meaning using a unimodal mass--observable relation equivalent to a $\gamma(z) = 0$ assumption in Eq.~\ref{eqn:blendgamma}, then the resultant projection-free counts, ${\overline N_1}$, and sample covariance $\mathbf{S_1}$ may differ from the true values of ${\overline N}$ and $\mathbf{S}$, respectively.  The set of model (cosmological + mass--observable) parameters, $\thetaCosm$, recovered  will generally differ from that of the true model.  
The bias in the model parameters is given by \citep{wu08},
\begin{multline}
\delta \theta_{\alpha} = \sum_{\beta} \left( F_1^{-1} \right)_{\alpha \beta} \Bigl[ ({\overline N} - {\overline N_1} )^{t} \mathbf{C_1}^{-1} {\overline N_1,_{\beta}} \\
+ \frac {1}{2} \mathrm{Tr} \left\{ \mathbf{C_1}^{-1} \mathbf{S_1},_{\beta} \mathbf{C_1}^{-1} \left( \mathbf{S} - \mathbf{S_1} \right) \right\} \Bigr].
\label{eqn:shifts}
\end{multline}

The covariance and Fisher matrix in the above expression are evaluated for the projection-free model using parameter values determined by fitting the redshift behavior of the first two moments of the mass-observable relation with projection.  For a specific choice of true model parameters  {$\mu_0$, $\alpha$, $\gamma_1$}, (and fixing $\gamma_0, \beta, s_z, s_M = 0$ and $\sigmaBlend^2 = 2\sigma_0^2$),  we compute the mean mass and variance in redshift bins of width $0.1$  
and fit these to determine the terms {$B_0$, $B_1$, $s_1$, $s_2$, $s_3$} of the unimodal model, Eqs.~(\ref{eqn:mbiasbl}) and (\ref{eqn:msigbl}).   Values for the case of $\gamma_1=0.15$ are given in Table \ref{tbl:fitparameters}.

\begin{table}[!ht]
\caption{Projection-free mass-observable parameters fit to the case with projection for $\gamma_1=0.15$.}
\label{tbl:fitparameters}
\begin{center}
\begin{tabular}{c c | c c c c r@{.}l r@{.}l}\hline \hline
$\mu_0$ & $\alpha$ & $B_0$ & $B_1$ & $\sigma_0^2$ & $s_1$ & \multicolumn{2}{c}{$s_2$} & \multicolumn{2}{c}{$s_3$} \\ \hline
0.75 & 0.00 & 0.0076 & 0.0389 & 0.2503 & 0.1110 & -0&1299 & 0&0491 \\
0.50 & 0.50 & 0.0040 & 0.0470 & 0.2500 & 0.0760 & -0&0497 & 0&0097 \\
0.25 & 1.00 & 0.0004 & 0.0551 & 0.2499 & 0.0475 & 0&0111 & -0&0163 \\
\hline
\end{tabular}
\end{center} 
\end{table}

For $\gamma_1=0.15$, the shifts in the mean mass are below one percent at $z=0$ but grow to $3.8\%$ at $z=1$ for the strong blending evolution case ($\alpha  = 1$).  The mass bias fit, constrained by the form of Eq.~\eqref{eqn:mbiasc} with only two free parameters, can differ from the true bias in the projection model by up to $0.007$ at $z=1$ when the fit is the worst (in the $\alpha=0$) case, but only by $0.002$ for the best fit ($\alpha = 1$) case.  
The redshift behavior of the variance, with four free parameters of Eq.~\eqref{eqn:msigc}, matches the values of the projection case quite well, with deviations less than $3 \times 10^{-4}$ in the worst case.
The variance at $z = 1$ is larger for larger values of $\alpha$, with $\sigma^2 = 0.293$ for the $\alpha = 1$ case.

For smaller $\gamma$ values, the fits deviate less from the true bias in projection.  For comparison, the bias at $z=1$ for a $\gamma_1=0.05$ fit, differs by $0.002$ in the worst case and the mass bias is about $0.03$.  The variance also grows more slowly with $z$, with $\sigma^2 = 0.264$ at $z=1$  for the $\alpha=1$ case.

We compute survey expectations for counts ($\overline{N}$) in mass, angle and redshift bins and their covariance ($\mathbf{S}$) for the range $0 \le \gamma_1 \le 0.3$.  We then calculate the counts ($\overline{N}_1$), sample covariance ($\mathbf{S_1}$), full covariance ($\mathbf{C_1}^{-1}$), and Fisher matrix ($F_1$) for the respective projection-free case using the best-fit parameters described above.  As mentioned in Sec. \ref{sec:clustermass}, we add unbiased \Planck\  priors to the Fisher matrix, so that $F_1 \rightarrow F_1+F_{\rm Planck}$.  The resultant values are used to compute bias in model parameters according to Eq.~\eqref{eqn:shifts}.

\psfrag{deltaw}{\Huge $\delta w$}
\psfrag{-deltaDE}{\Huge $-\delta \DE$}
\psfrag{dth}[c][c][1.25][90]{\Huge $\delta \theta$}
\psfrag{gamma}[c][c][1.25][0]{\Huge $\gamma$}
\psfrag{munot = 0.25, alpha = 1.00}{\Huge $\mu_0 = 0.25, \alpha = 1.0$}
\psfrag{munot = 0.50, alpha = 0.50}{\Huge $\mu_0 = 0.50, \alpha = 0.5$}
\psfrag{munot = 0.75, alpha = 0.00}{\Huge $\mu_0 = 0.75, \alpha = 0$}
\psfrag{z}{\Huge $z$}
\psfrag{f}{\Huge $f_{\rm blend}$}
\psfrag{0}{\Huge $0$}
\psfrag{0.05}{\Huge $0.05$}
\psfrag{0.1}{\Huge $0.10$}
\psfrag{0.15}{\Huge $0.15$}
\psfrag{0.2}{\Huge $0.20$}
\psfrag{0.25}{\Huge $0.25$}
\psfrag{0.2}{\Huge $0.2$}
\psfrag{0.3}{\Huge $0.3$}
\psfrag{1}{\Huge $1$}
\begin{figure}[!t]
  \begin{minipage}[h]{80mm}
    \begin{center}
      \resizebox{80mm}{!}{\includegraphics{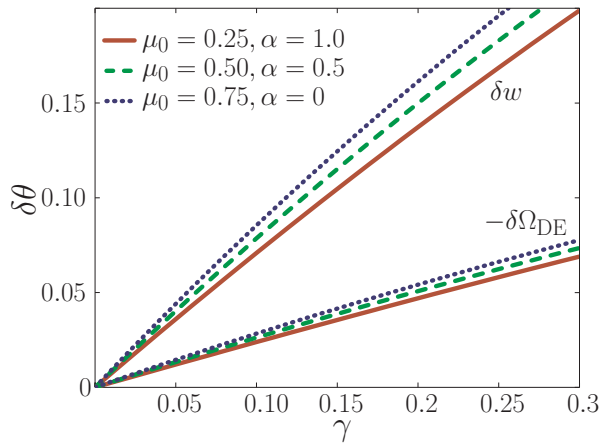}}
    \end{center}
  \end{minipage}
  \caption{Shifts in the cosmological parameters $w$ (upper lines) and $-\DE$ (lower lines) as a function of the blending evolution rate, $\gamma_1$.   Line-styles correspond to three different redshift dependence forms (none, weak, strong) shown in Fig.~\ref{fig:proj}.  
  }
  \label{fig:deltas}
\end{figure}

Fig.~\ref{fig:deltas} shows the resulting biases in $w$ and $\DE$.  For the cases shown, we assume a \Planck\  prior on the cosmological parameters but all other model parameters are free.
When $\gamma_1=0$ there is no blended component and therefore no parameter bias (note we assume $\gamma_0 = 0$).  The bias in cosmological parameter estimates grows approximately linearly with $\gamma_1$, with slopes that are weakly dependent on the assumed redshift evolution of the mass bias in the blended component.  For the strong redshift evolution case ($\mu_0 = 0.25, \alpha = 1.0$) with $\gamma_1=0.17$,  which provides a close match to the simulation results of \citet{coh07}, we find a significant biases in the dark energy equation of state, $\delta w = 0.12$, and in the dark energy density, $\delta \DE = -0.04$.  

These shifts may be considered pessimistic, in the sense that we have assumed a large scatter in the blended component.  For the case of $\sigma^2_{\rm blend} = \sigma^2_{0}$, the slopes of the equivalent lines in Fig.~\ref{fig:deltas} are reduced by $\sim 50\%$, so that the strong redshift evolution case with $\gamma_1=0.17$ produces $\delta w = 0.08$ and $\delta \DE = -0.03$. Reducing the assumed $\sigma_0 = 0.5$ scatter in the clean component would also lead to smaller biases in cosmological parameters.  

\begin{table}[!ht]
\caption{Cosmological parameter shifts, $\delta \theta$, for strong redshift evolution ($\mu_0=0, \alpha = 1.0$) and $\gamma_1=0.17$.}
\label{tbl:shifts}
\begin{center}
\begin{tabular}{c r@{.}l r@{.}l }\hline \hline
Parameter & \multicolumn{2}{c}{$\theta_{\rm true}$} & \multicolumn{2}{c}{$\delta \theta$} \\ \hline
$\Omega_b h^2$ & 0&0227 & -0&0001 \\
$\Omega_m h^2$ & 0&1326 & 0&0009 \\
$\DE$ & 0&742 & -0&0401 \\
$w$ & -1&0 & 0&1178 \\
$\delta_{\zeta} \times 10^5$ & 4&625 & 0&0222 \\
$n$ & 0&963 & -0&0015 \\
$\tau$ & 0&087 & \multicolumn{2}{c}{$1.0 \times 10^{-7}$} \\ 
\hline
\end{tabular}
\end{center} 
\end{table}

Table \ref{tbl:shifts} shows the bias in all cosmological parameters for strong redshift evolution for $\gamma_1=0.17$, the case that best matches \citet{coh07}.
The bias for parameters other than $\DE$ and $w$ is less than 1\% of the fiducial value.
However, comparing to the fiducial uncertainties from the Fisher matrix with unbiased \Planck\  priors show that the shifts can approach a 1-$\sigma$ level for $\Omega_m h^2$ and $\delta_{\zeta}$. 

\begin{figure}[!t]
  \begin{minipage}[h]{95mm}
    \begin{center}
      \resizebox{95mm}{!}{\includegraphics{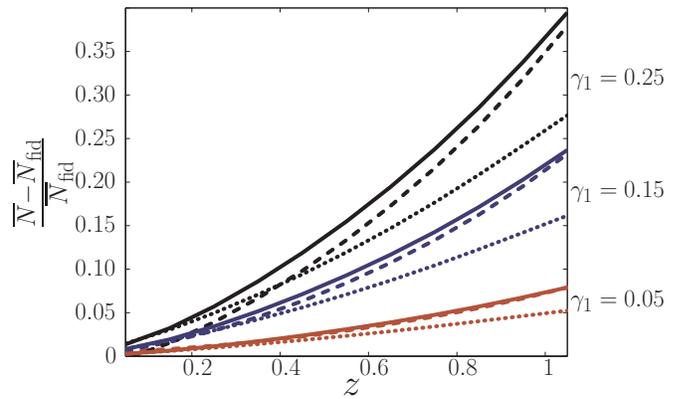}}
    \end{center}
 \end{minipage}
  \caption{Fractional change in counts in the strong redshift evolution case ($\mu_0 = 0.25, \alpha = 1$), relative to a projection-free model, are shown for three values of the blending evolution parameter, $\gamma_1$.  Solid lines give the case with projection while dotted lines show the projection-free model with parameters tuned to match the mass bias and variance of the projection model, but with cosmology fixed at the fiducial WMAP5 values.  Dashed lines show the projection-free case after shifting all parameters (cosmological and mass-observable)  according to Eq.~\eqref{eqn:shifts}.}
  \label{fig:shiftedcounts}  
  \end{figure}

Fig.~\ref{fig:shiftedcounts} offers insight into the magnitude of the change in cluster counts arising from projection.  As a fiducial measure, we use counts, $\overline{N}_{\rm fid}$, for the projection-free (unimodal) case with default parameters (zero bias and redshift-independent variance).  The solid lines in Fig. \ref{fig:shiftedcounts} show the fractional shifts in counts, relative to the fiducial, as a function of redshift for the projected (bimodal) cases with $\mu_0=0.25, \alpha = 1.0$.   For $\gamma_1 \gta 0.1$, projection boosts counts on the order of a few tens of percent at high redshift.    The dotted lines show projection-free expectations when the mass--observable parameters are shifted to the values given in Table \ref{tbl:fitparameters}, but the cosmology is held fixed.   The dashed lines give projection-free expectations when both cosmological and mass--observable parameters are adjusted according to Eq.~\eqref{eqn:shifts}.

\psfrag{Nbarp}[c][c][1.25][90]{\Huge $\frac{\overline{N} - \overline{N}_{\rm fid}}{\overline{N}_{\rm fid}}$}
\psfrag{z}[c][c][1.5][0]{\Huge $z$}
\psfrag{0}{\Huge $0$}
\psfrag{0.30}{\Huge $0.30$}
\psfrag{0.35}{\Huge $0.35$}
\psfrag{0.40}{\Huge $0.40$}

The counts of the projection-free model with fully shifted parameters provide a good match to the counts with projection.  The adjustment of the mass--observable parameters alone offers a good match at low redshifts, but at high redshift, a unimodal fit to the bimodal form of the projected $p(\Mobs | M, z)$ becomes increasingly less accurate.  Adjustments in cosmological parameters shift the amplitude and shape of the mass function as a function of redshift, providing a degree of compensation for deficiencies introduced by a unimodal $p(\Mobs | M, z)$ assumption.  While the quality decreases for higher values of $\gamma_1$, the fits are still acceptable in a $\chi^2$ sense. 

Note that as $\gamma_1$ grows and the associated shifts in parameters grow, the linear approximation for the bias given by Eq.~\eqref{eqn:shifts} begins to break down.  For $\gamma_1=0.05$ agreement between the shifted single-Gaussian case and the two-Gaussian case is quite good, while at $\gamma_1=0.25$ the divergence is much larger. 

Finally, we note that Eq.~\eqref{eqn:shifts} calculates shifts using the Fisher matrix of the projection-free model.  We have verified that we obtain the same results if we employ the projection model matrix with sharp priors added to the eight parameters describing the blended component.  
This is expected because, for the same free parameters in the two models, the linearized equation should be symmetric under their exchange.

\subsection{Dark Energy Parameter Constraints}\label{ssec:fidcons}

While introducing additional parameters to describe selection with projected blending may eliminate bias in cosmological parameters, that benefit comes with the risk of degrading cosmological parameter constraints.  The amount of degradation depends on assumptions about priors on model parameters.   

Table~\ref{tbl:constraints} summarizes results using the projection model that 
corresponds to our best match of \citet{coh07} ($\mu_0 = 0.25,\ \alpha = 1.00$, $\beta, s_z, s_M = 0$, $\sigmaBlend^2 = 2 \sigma_0^2$, $\gamma_0=0,\ \gamma_1 = 0.17$).    In all cases, \Planck\  priors are added to the cosmological parameters, and we consider priors on the clean and blended cluster components separately.  Given an assumed prior error, $\sigma_i$, on the $i^{\rm th}$ parameter, we add to the Fisher matrix
\begin{equation}
F_{\rm prior}^{ii} = \left( \frac {1}{\sigma_i} \right)^2.
\label{eq:pri}
\end{equation}
We consider sharp priors as being numerically larger than other entries in the Fisher matrix, generally $F_{\rm sharp} \simeq 10^6$, and flat priors are given by $F_{\rm flat} = 0$.

\begin{table}[!ht]
\caption{Forecasts for $w$ and $\DE$ constraints using \Planck\  priors.}
\label{tbl:constraints}
\begin{center}
\begin{tabular}{ l l | c c }\hline \hline 
 \multicolumn{2}{c |}{Priors} & \multicolumn{2}{| c}{Uncertainty} \\ \hline
Clean  & Blended & $\sigma(\DE)$ & $\sigma(w)$ \\ \hline
sharp & sharp & 0.002 & 0.011 \\ 
flat & sharp & 0.014 & 0.046 \\ 
flat & flat & 0.034 & 0.109 \\
\hline
0.1 & sharp & 0.010 & 0.030 \\
0.1 & 0.1 & 0.010 & 0.030 \\
0.1 & flat & 0.013 & 0.042 \\
\hline
\end{tabular}
\end{center} 
\end{table}

Table~\ref{tbl:constraints} shows permutations of three basic cases: a flat prior on model parameters,  a prior of $\sigma = 0.1$ added to model parameters as well as a 10\% prior added to $\sigma_0^2$ or $\sigmaBlend^2$, or sharp priors on model parameters.   The last two columns give the marginalized uncertainty in $w$ and $\DE$.

The first three rows compare extremal cases.   Sharp knowledge of all mass--observable parameters produces the best constraints possible, $\pm 0.002$ in $\DE$  and $\pm 0.011$ in $w$.   The projection--free case with no prior knowledge of the six parameters of the clean component, shown in the second row, produces constraints of $\pm 0.014$ in $\DE$  and $\pm 0.046$ in $w$.  These errors are worse by factors of $7$ and $4$, respectively, than the case of perfect knowledge.  Introducing eight new degrees of freedom to represent the blended component further degrades the errors by somewhat more than a factor of two, to $\pm 0.034$ in $\DE$  and $\pm 0.11$ in $w$.
 
Targeted follow-up and complementary survey information, from mm or X-ray observations for example, may enable moderate priors to be placed on the bias and variance of the mass--observable relation.  These cases are explored in the lower three rows of Table~\ref{tbl:constraints}.   Knowledge of the clean component parameters at the level of $\pm 0.1$ provides substantial improvement over the flat case.  Even with no prior knowledge of the blended component, the errors of $\pm 0.013$ in $\DE$  and $\pm 0.042$ in $w$ represent improvements over the projection--free case with no prior knowledge (second row).   When $0.1$ priors are brought to bear on the projected blends, the constraints improve to $\pm 0.010$ in $\DE$  and $\pm 0.030$ in $w$.  Stronger priors on the blended component do not improve these constraints.  

\subsection{Discussion}\label{ssec:disc}

Achieving constraints on $w$ and $\DE$ at the few percent level is a goal of next-generation cluster surveys.  Our analysis shows that avoiding biases at this level requires projection to be incorporated into the likelihood modeling of optical-IR surveys.

\begin{figure*}[!ht]
  \begin{minipage}[h]{85mm}
    \begin{center}
      \resizebox{85mm}{!}{\includegraphics[angle=0]{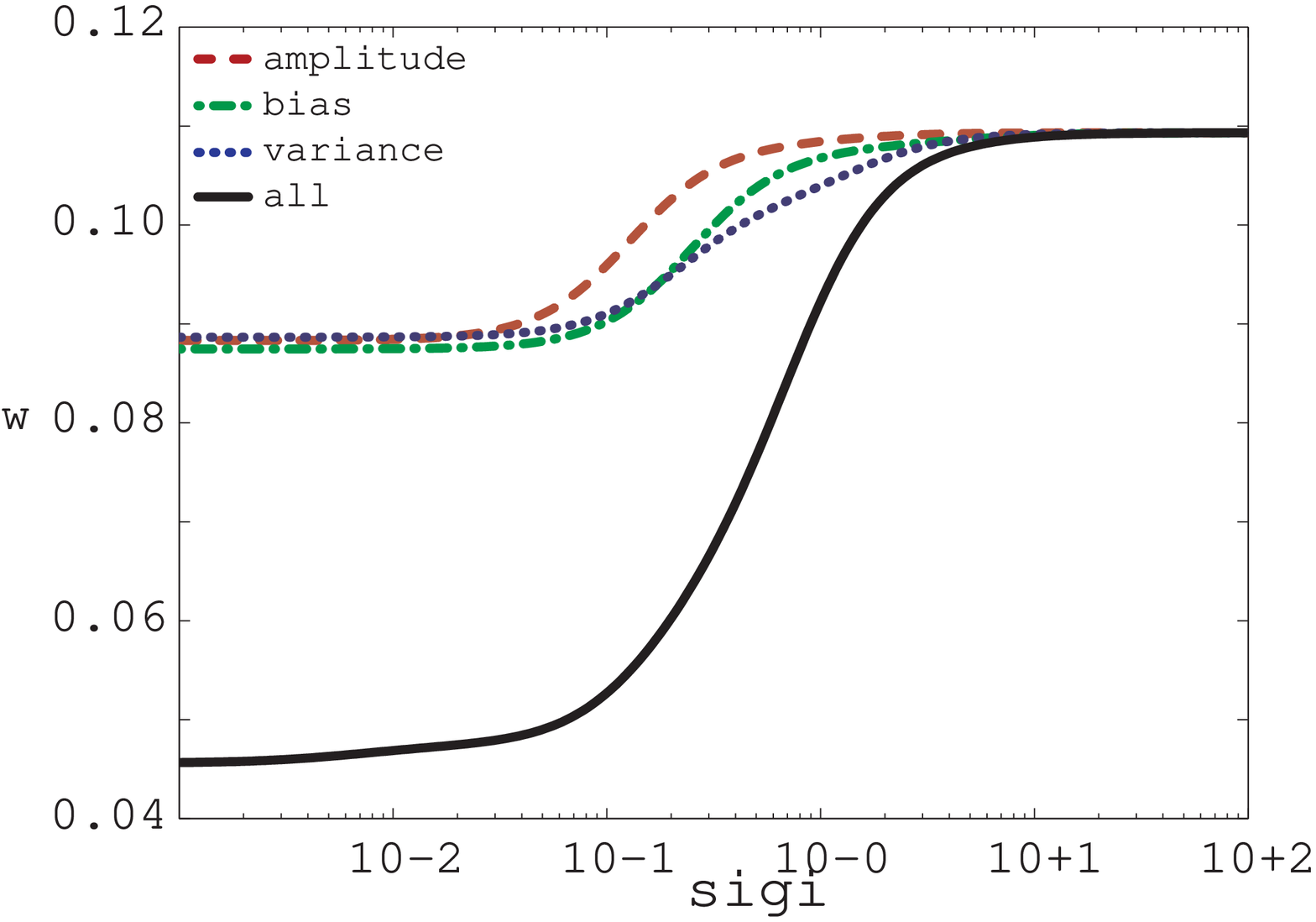}}
    \end{center}
  \end{minipage}
    \begin{minipage}[h]{85mm}
    \begin{center}
      \resizebox{85mm}{!}{\includegraphics[angle=0]{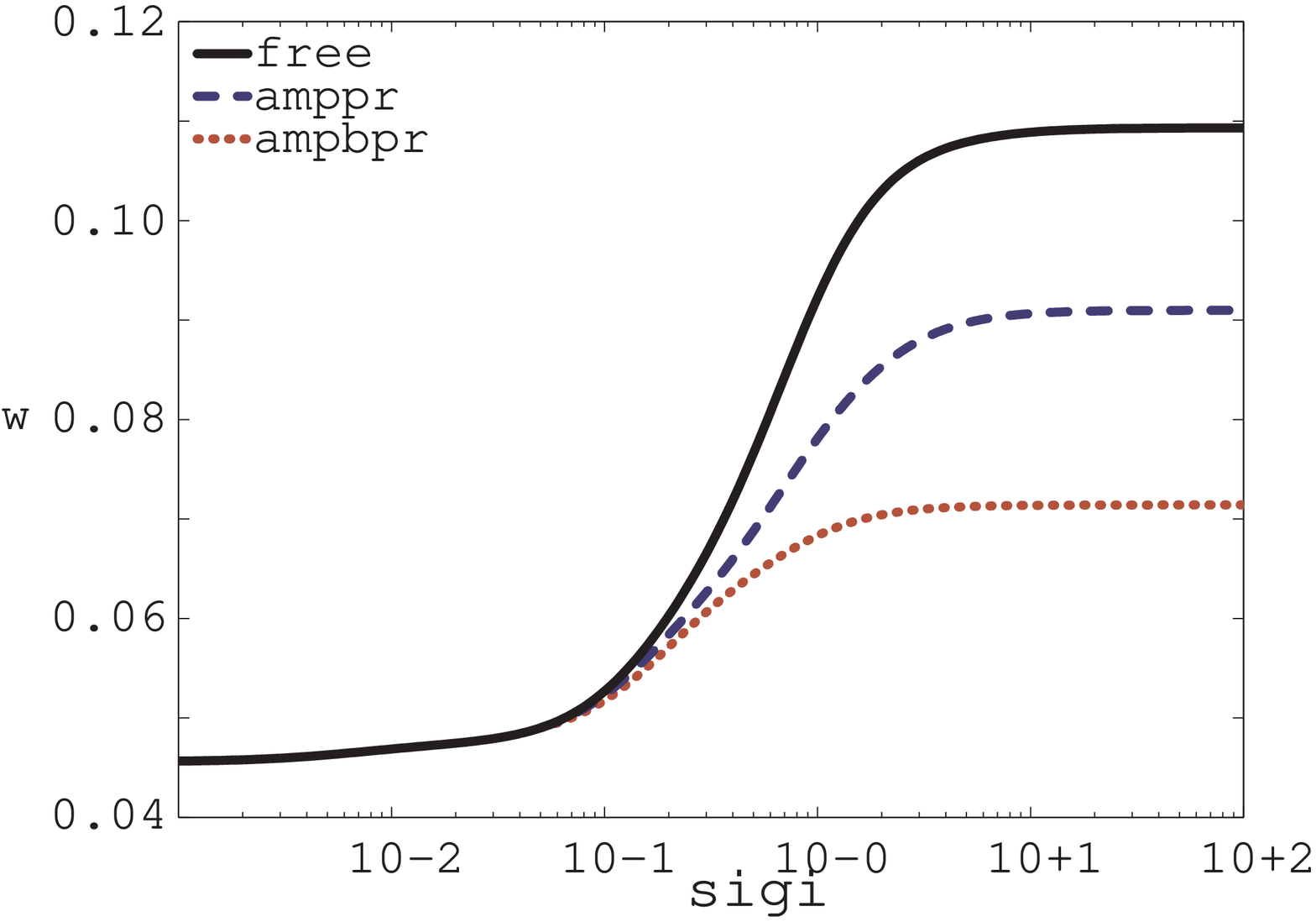}}
    \end{center}
  \end{minipage}
  \caption{Error in $w$ is shown as a function of priors added to the parameters describing projected blends.  \Planck\ priors are applied to cosmological parameters throughout. The {\it left} panel shows the error in $w$ as a function of prior $\sigma_i$ applied to all eight blending parameters (black solid), and separately to only the parameters describing $\gamma(z)$ (red dashed), or only the parameters describing ${\rm ln} M_{\rm bias,b}$ (green dot-dashed), or only the parameters for $\sigma^2_{{\rm ln}M,{\rm b}}$ (blue dotted).  The {\it solid black} line is reproduced in the {\it right} panel, which also shows the resulting error in $w$ when a $0.05$ prior is applied to just the blending amplitude, $\gamma(z)$ (blue dashed), or when $0.05$ priors are applied to both $\gamma(z)$ and ${\rm ln} M_{\rm bias,b}$ (red dotted).  In the latter two cases, the results are shown as a function of prior $\sigma_i$ applied to the remaining blending parameters.   
  } \label{fig:deg}
\end{figure*}

Prior knowledge of the blended component behavior can limit parameter bias.  In Fig.~\ref{fig:deg}, we illustrate the rate at which the forecast uncertainty in $w$ changes with prior uncertainty on the mass--observable parameters of the blended component.  The behavior for $\DE$ is similar, mainly because \Planck\ priors effectively fix many of the correlations between cosmological parameters.  The solid line shows the effect of applying priors to all eight parameters while dashed lines show the behavior when priors are applied only to parameters controlling the blending amplitude ($\gamma_0, \ \gamma_1$), mass bias terms ($\mu_0, \ \alpha, \  \beta$) and mass variance ($\siglnMBlend , \ s_z, \ s_M$).   In all cases, flat priors are imposed to the remaining mass--observable parameters.  

\psfrag{amplitude}{\Huge amplitude}
\psfrag{bias}{\Huge bias}
\psfrag{variance}{\Huge variance}
\psfrag{free}{\Huge free}
\psfrag{all}{\Huge all}
\psfrag{amppr}{\Huge amplitude prior}
\psfrag{ampbpr}{\Huge amplitude and bias prior}
\psfrag{w}[c][c][1.25][90]{\Huge $\sigma(w)$}
\psfrag{sigi}[c][c][1.25][0]{\Huge $\sigma_i$}
\psfrag{z}{\Huge $z$}
\psfrag{f}{\Huge $f_{\rm blend}$}
\psfrag{0}{\Huge $0$}
\psfrag{0.04}{\Huge $0.04$}
\psfrag{0.05}{\Huge $0.05$}
\psfrag{0.06}{\Huge $0.06$}
\psfrag{0.07}{\Huge $0.07$}
\psfrag{0.08}{\Huge $0.08$}
\psfrag{0.09}{\Huge $0.09$}
\psfrag{0.10}{\Huge $0.10$}
\psfrag{0.11}{\Huge $0.11$}
\psfrag{0.12}{\Huge $0.12$}
\psfrag{10-2}{\Huge $10^{-2}$}
\psfrag{10-1}{\Huge $10^{-1}$}
\psfrag{10-0}{\Huge $10^{0}$}
\psfrag{10+1}{\Huge $10^{1}$}
\psfrag{10+2}{\Huge $10^{2}$}

Applying priors to only the parameter subsets, we see that all three sets have comparable effects on $w$ uncertainties.  Because of covariance, the effect of applying strong priors to all parameters is much stronger than for any of the isolated sets.  The error in $w$ is somewhat more sensitive to the blending amplitude $\gamma(z)$ than to the bias and variance, but all parameters need to be known at the level of $0.1$ in order to avoid significant degradation.   

For red sequence or photometric redshift cluster finding methods, the fraction of blended clusters is not likely to dominate the population, suggesting that the current level of uncertainty is near $0.1$.   By testing the performance of cluster finding algorithms on sophisticated simulations, and by calibrating mass selection based on multi-wavelength follow-up campaigns of existing deep cluster catalogs, the error may be reduced to $\sim 0.05$ as part of next generation survey analysis.  In the right-hand panel of Fig.~\ref{fig:deg}, we show the error in $w$ for cases in which just the amplitude or both the amplitude and mass bias of the blending component are known at the $\sim 0.05$ level.   These cases limit the degradation of the $w$ constraint, to factors of $2.0$ and $1.5$, respectively, compared to $2.4$ for the case of all parameters free. 

Our assumed value of $\sigma_0 = 0.5$ may be pessimistic, in that future cluster finders may achieve better mass selection.  An improved matched filter method \cite{Rykoff1104.2089} applied to the \maxbcg\ catalog indicates a mass scatter closer to $0.3$ for low redshift clusters for a sample with mass threshold close to the value assumed here.   While achieving this level of mass selection at $z > 0.5$ has not yet been demonstrated, the variety of cluster multi-color detection algorithms under active development \cite{Miller0503713, Koester0701265, Dong0709.0759, Wen0906.0803, Adami0910.3827, milk10, Soares-Santos1011.3458, Song1104.2332} offer the potential of future gains.  

Spectroscopic observations of cluster fields provide valuable empirical tests of blending.  For example, a follow-up of 58 EIS cluster candidates, selected only with I-band imaging and so without the benefit of color-based redshift filtering, found multiple redshift-space structures in a majority of fields \cite{Grove0809.4552}.   Studies at high redshift using infrared color selection, which are just beginning, fare better but are not entirely clean.  Six rich clusters from the SPARCS sample, which uses $z^\prime$--3.6$\mu$ color from ground-based and {\sl Spitzer} observations, have been followed up with Keck/LRIS spectroscopy.   Two cases appear to be strongly blended, with dynamical mass estimates derived from velocity dispersions lower by a factor $\sim 6$ than mass estimates based on their galactic richness.  Continued follow-up of this and other IR-selected deep cluster samples should be followed vigorously as a means to characterize the amplitude and mass scale of projected blends.  

Simulations of large-scale structure provide an effective tool for understanding projection.  Work is underway within the DES collaboration to test a variety of cluster finding algorithms against simulated expectations for the multi-band galaxy catalog.   Using either galaxy membership or redshift-space location as a way to match clusters and halos, the simulations offer the means to test the sensitivity of blending to algorithm choice and to choice of parameters within a fixed algorithm \cite{Gerke0410721}.  Such studies should produce improved, algorithm-specific characterizations of blends that can be coupled to empirically-derived studies to serve as prior information for cluster likelihood analysis. 

As algorithms improve in terms of mass selection, characterization of projection effects will inevitably become apparent through the full shape of the mass--observable relation, $p(\Mobs | M,z)$ or its inverse, the mass--selection function, $p(M | \Mobs,z)$.  Ultimately, survey constraints on dark energy parameters have the potential to achieve the best possible constraints given by the first row of Table~\ref{tbl:constraints}.   Extracting a one percent constraint on $w$ poses the challenge of precisely characterizing selection.  More careful analysis may suggest an improved, possibly more compact, form for modeling selection with projection than what we present here.  

While we focus our analysis on optical-IR studies, the issue of blending is generic to all cluster finding methods.  The blending factor $\gamma(z)$ should be minimal for X-ray selection, due to the compactness of the surface brightness image as well as the strong scaling of luminosity with mass.  SZ selected samples are likely to incur blending at a level below that of optical-IR surveys \cite{coh09}.  However, for X-ray and SZ, angular resolution is also an important factor.   The \Planck\ satellite has only moderate resolution of 5--10$^\prime$, depending on frequency.   Of the 21 new cluster candidates identified in the \Planck\ ESZ sample, four are known to be double or triple systems from XMM follow-up imaging \cite{PlanckXMMfollowup11}.  Follow-up studies of these and other SZ-selected sources from SPT and ACT is needed to characterize the mass selection of these methods.  

\section{Summary}\label{sec:conc}

Cluster counts used in cosmological studies have typically been modeled with log-normal deviations about power-law forms for the mass--observable relation.  While a log-normal expectation may reasonably reflect intrinsic scatter, projection will generically boost a minority of systems to higher observed signal.  This blending of halo properties is particularly true for optical-IR surveys that use color or photometric redshifts as a distance estimator.  We extend previous Fisher matrix studies by introducing a Gaussian mixture model for the mass--observable relation.  The model employs eight new parameters to describe a redshift-dependent amplitude and shape of the blended component, in addition to the six parameters of the dominant, non-blended cluster population. 

The presence of a minority of strongly blended clusters influences cosmological parameter estimation.  
For the case of blending parameters tuned to Millennium Simulation analysis \cite{coh07} (Fig.~\ref{fig:proj}), we find that survey analysis using a projection--free (single Gaussian) analysis model introduces biases of $0.1$ in $w$ and $-0.04$ in $\DE$.  Comparing their Fisher forecast errors with \Planck\ , these shifts are comparable to uncertainties expected using flat priors on mass--observable parameters, but are an order of magnitude larger than the uncertainties possible under precise mass--observable knowledge.  Explicit modeling of projection is therefore required to avoid significant bias in next generation cosmological studies using cluster counts and clustering.  Optical studies at low redshift, where the blended fraction is below ten percent, or studies using cleaner detection methods, such as X-ray selection, are less susceptible to bias from projection.  

Constraints on $w$ and $\DE$ with \Planck\  priors degrade by about a factor of $2.4$ when new parameters to describe the Gaussian mixture distribution are included.   Our analysis indicates that $5\%$ prior knowledge of the blending amplitude and mass bias limit the degradation to a factor of $1.5$.   

Improved knowledge of blending will come from complementary approaches employing follow-up observations, simulations, and joint analysis of overlapping multi-wavelength surveys.  Follow-up campaigns will provide mass estimates based on hydrostatic, virial and lensing masses.  Simulations of the galaxy distribution will grow in fidelity, benefitting from empirical studies of the relation between halo mass stellar content to $z \sim 1$ \cite{Leauthaud1103.2077, Leauthaud1104.0928}.   Optical cluster finders applied to such simulated sky expectations will inform prior constraints on projection effects.   Over the next decade, the ability to cross-match large cluster samples from mm to X-ray wavelengths will offer a new window into the nature of the relationship between clusters and the massive halos that host them.  

\appendix
\section{Cluster Counts and Clustering}\label{sec:countsandclustering}

For the space density and clustering of halos, we follow conventions used in previous work \cite{cunevr10}.  The mass function is 
\begin{equation}
\frac{dn}{dM} = f(\sigma)\frac{\overline \rho_m}{M}\frac{d{\rm ln} \sigma^{-1}}{dM} , \label{eqn:mfunc}
\end{equation}

\noindent and we adopt the Tinker parameterization of $f(\sigma)$ \citep{Tinker0803.2706}, 
\begin{equation}
f(\sigma) = A\left[\left(\frac{\sigma}{b}\right)^{-a} +1 \right]e^{-c/{\sigma^2}}. \label{eqn:tinker}
\end{equation}

For fiducial parameters, we adopt the values of \cite{Tinker0803.2706} at mean density contrast $\Delta=200$:
$A_0 = 0.186$, $A_x =-0.14$, $a_0 = 1.47$, $a_x = -0.06$, $b_0 = 2.57$, $\log_{10}(\alpha) = {(\frac{0.75}{log(\Delta/75)})^{1.2}}$, and $c = 1.19$.


The sample covariance of counts $N_{\alpha,i}$ is, given by \citep{hu03}

\begin{eqnarray}
S^{\alpha \beta}_{ij} &=& \langle (N_{\alpha,i} -\overline N_{\alpha,i})(N_{\beta,j} - \overline N_{\beta,j})\rangle \label{eqn:sija} \\
&=& b_{\alpha,i} \overline N_{\alpha,i} b_{\beta,j} \overline N_{\beta,j} \nonumber \\
&& \times \int{\frac{d^3 k}{(2\pi)^3}} W_i^*({\bf k})W_j({\bf k})\sqrt{P_i(k)P_j(k)}, \label{eqn:sijb}
\end{eqnarray}

\noindent where $b_{\alpha,i}(z)$ is the 
average cluster linear bias parameter, defined as
\begin{multline}
b_{\alpha,i}(z) = \frac{1}{\overline n_{\alpha,i}(z)}  \int \frac{d{\Mobs^\alpha}}{\Mobs^\alpha}\int \frac{d{\Mobs^\beta}}{\Mobs^\beta} \int \frac{d M}{M} \\
\times \frac{d \overline n_{\alpha,i}(z)}{d{\rm ln} M} b(M;z)p(\Mobs|M).
\end{multline}

\noindent $W_i^*({\bf k})$ is the Fourier transform of the top-hat window function and $P_i(k)$ is the linear power spectrum at the centroid of redshift bin $i$.  

We adopt the $b(M,z)$ fit of \cite{she99} for the halo bias
\begin{equation}
b(M,z) = 1 + \frac{a_c \delta_c^2/\sigma^2 -1}{\delta_c} 
         + \frac{ 2 p_c}{\delta_c [ 1 + (a \delta_c^2/\sigma^2)^{p_c}]}\label{eqn:biasofmz},
\end{equation}
\noindent and choose the fiducial values for the parameters to be
$a_c=0.75$, $\delta_c=1.69$, and $p_c= 0.3$.

Following \cite{hu03}, we find that the window function  $W_i^*({\bf k})$ is given by

\begin{equation}
W_i({\bf k}) = 2\exp{\left[ i k_{\parallel} \left( r_i \right) \right] }
            \frac{\sin( k_{\parallel} \delta r_i/2) }{k_{\parallel} \delta r_i/2} 
              \frac{J_1(k_{\perp} r_i \theta_s)}{k_{\perp} r_i \theta_s}. 
\label{eqn:window}
\end{equation}
\noindent Here $r_i=r(z_i)$ is the angular diameter distance to the 
$i^{\rm th}$ redshift bin, and $\delta r_i=r(z_{i+1})-r(z_{i})$.
Similarly, $H_i=H(z_i)=H(z)$, which we assume to be constant inside each bin.
The variables $k_{\parallel}$ and $k_{\perp}$ represent parallel and perpendicular components of the wavenumber ${\bf k}$ relative to the line of sight.

\acknowledgments

We are grateful to Dragan Huterer, Risa Wechsler and Heidi Wu for valuable conversations, and to Wayne Hu for providing the \Planck\  prior matrix used in our analysis.  We also thank Laurie Shaw and Gil Holder for reviewing a draft. AEE acknowledges support from NSF AST-0708150 and NASA NNX07AN58G. CC is supported by the DOE OJI grant under contract DE-FG02-95ER40899.

\bibliography{brandon}
\end{document}